\documentclass[conference,harvard,brazil,english,portuguese]{sbatex}

\makeatletter
\def\verbatim@font{\normalfont\ttfamily\footnotesize}
\makeatother

\usepackage[utf8]{inputenc}

\usepackage{icomma} 
\usepackage{amsmath,amssymb,amsfonts} 
\usepackage{verbatim}
\usepackage{graphicx} 
\usepackage{psfrag} 
\usepackage{float} 
\usepackage{multirow}
\usepackage{multicol, blindtext, graphicx}
\usepackage{color}
\usepackage{epstopdf}
\usepackage{xcolor}
\newcommand\crule[3][lightgray]{\textcolor{#1}{\rule{#2}{#3}}}
\usepackage{hyperref}
\usepackage{url} 

\newtheorem{theo}{Teorema}

\begin{document}

\title{IDENTIFICA\c{C}\~AO DE SISTEMAS DIN\^AMICOS COM ARITM\'ETICA INTERVALAR}
\author{Márcia L. C. Peixoto}{marciapeixoto93@hotmail.com}
\address{Programa de Pós-Graduação em Engenharia Elétrica (UFSJ/CEFET)\\  UFSJ - Universidade Federal de São João del-Rei\\ Pça. Frei Orlando, 170 - Centro - 36307-352 - São João del-Rei, MG, Brasil}
\address{GCOM - Grupo de Controle e Modelagem \\ UFSJ - Universidade Federal de São João del-Rei\\ Pça. Frei Orlando, 170 - Centro - 36307-352 - São João del-Rei, MG, Brasil}

\author[1]{Marco T. R. Matos}{mtrmatos@hotmail.com}
\author[2]{Wilson R. Lacerda Júnior}{wilsonrljr@outlook.com}
\author[2]{Samir A. M. Martins}{martins@ufsj.edu.br}
\author[1]{Erivelton G. Nepomuceno}{nepomuceno@ufsj.edu.br}

\twocolumn[

\maketitle


\selectlanguage{english}
\begin{abstract}
This paper aims to identify three electrical systems: a series RLC circuit, a motor/ generator coupled system, and the Duffing-Ueda oscillator. In order to obtain the system's models was used the error reduction ratio and the Akaike information criterion. Our approach to handle the numerical errors was the interval arithmetic by means of the resolution of the least squares estimation. The routines was implemented in Intlab, a Matlab toolbox devoted to arithmetic interval.  Finally, the interval RMSE was calculated to verify the quality of the obtained models. The applied methodology was satisfactory, since the obtained intervals encompass the system's data and allow to demonstrate how the numerical errors affect the answers.

\end{abstract}


\selectlanguage{brazil}
\begin{abstract}
Neste trabalho é feita a identificação de três sistemas: um circuito RLC série, um sistema composto por um motor e gerador acoplados  e o oscilador {\it Duffing-Ueda}. Foram utilizados a taxa de redução de erro e o critério de Akaike para a obtenção dos modelos dos sistemas. Com o intuito de manusear os erros numéricos aplicou-se a aritmética intervalar na resolução dos mínimos quadrados,  por meio do \textit{toolbox} Intlab. Por fim, foi calculado o índice RMSE intervalar com o intuito de verificar a qualidade dos modelos obtidos. A metodologia aplicada foi satisfatória visto que, os intervalos obtidos englobam os dados do sistemas e ainda sendo possível observar como os erros numéricos afetam as respostas. 
\end{abstract}

]

\selectlanguage{brazil}

\section{Introdução}

A Identificação de Sistemas estuda maneiras de modelar e analisar sistemas na tentativa de encontrar algum padrão em observações \cite{Billings2013,aguirre2015}. Para identificar um sistema, é necessário a suposição de um modelo que seja capaz de descrever as características lineares e não-lineares do mesmo.  Define-se modelo matemático como uma aproximação, constituída por equações que representam um sistema real. Estas equações quantificam as relações existentes entre as variáveis manipuladas com as variáveis controladas. 
Na literatura há diversas formas para a identificação de um mesmo sistema \cite{NTA2007}. São utilizadas várias representações matemáticas e computacionais, entre elas, pode-se citar as Redes Neurais, Lógica {\it Fuzzy}, Modelos NARMAX {\it(Nonlinear AutoRegressive Moving Average model with eXogenous input)} polinomiais, Modelo Baseado em Indivíduos, entre outros. 

Em Engenharia, essas técnicas podem ser empregadas na identificação de vários sistemas, como na identificação de aquecedores elétricos, conversores, sistemas químicos, entre outras áreas, como sistemas biológicos, econômicos e outros ramos da ciência \cite{MNF2011}.
Partindo de uma estrutura adequada, pode-se afirmar que  na identificação de sistemas existem principalmente duas fontes de erros: erros devido a incertezas de medições que são constituídos por ruídos inerentes ao sistema, imprecisão dos aparelhos utilizados na medição e ainda inexperiência na utilização dos aparelhos de medição \cite{Laguna2014}. E, a outra fonte de erro é devido as incertezas paramétricas existentes no sistema físico \cite{Peixoto2016}. 

Na identificação de sistemas é  importante o uso de simulações computacionais. Entretanto, simulações computacionais estão sujeitas a erros, mesmo que os dados de entrada sejam representáveis, o resultado de uma simples operação matemática pode não ser representável.
Em vez de um resultado verdadeiro, o computador retorna uma aproximação. Pequenos erros de arredondamento se acumulam e são propagados em sucessivos cálculos \cite{Nepomuceno2014,Galias2013,Ove2001}. \citeasnoun{Nepomuceno2016} mostram que soluções de modelos NARMAX que são matematicamente equivalentes, na solução computacional as respostas podem não ser equivalentes.

Assim, com a existência dos erros numéricos bem como com os erros intrínsecos presentes nos sistemas, a utilização da aritmética intervalar \cite{Moore1979,IEEE2015} tem sido considerada como um método eficiente para tratar destes erros. Métodos intervalares foram desenvolvidos com o objetivo de controlar erros de arredondamento em cálculos de ponto flutuante e seu uso cresce devido à motivação de se controlar estes erros \cite{Ruetsch2005,Moore1979}. A ideia é que em vez de usar um único valor de ponto flutuante para representar um número, o que implicaria em um erro se o número não é representável na máquina, o valor é representado por limites superior e inferior, os quais definem um intervalo representável na máquina. Atualmente a aritmética intervalar é aplicada em várias áreas, dentre sistemas elétricos de potência \cite{BSF+1994},  controle \cite{Banerjee2015}, processamentos de sinais \cite{Santana2012}, identificação de sistemas \cite{Laguna2014,Das}.

Neste trabalho é proposto a identificação de sistemas com enfoque na estimação dos parâmetros de modelos NARMAX com a contribuição da aritmética intervalar. O método é testado em três sistemas: um circuito RLC, um sistema eletromecânico, composto por um motor e gerador, e o oscilador {\it Duffing-Ueda}. Os dois primeiros sistemas foram excitados por um sinal binário pseudo-aleatório em suas respectivas entradas para a etapa de coleta de dados. Para seleção da estrutura foi utilizada a taxa de redução de erro juntamente com o critério de informação de Akaike \cite{Aka1974}. A aritmética intervalar foi aplicada nos mínimos quadrados, com o intuito de manusear os erros numéricos que são propagados durantes os sucessivos cálculos. E, por fim foi calculado o RMSE (\textit{Root Mean Square Error}), com o intuito de verificar a qualidade dos modelos obtidos.

O restante do trabalho está organizado da seguinte forma. Na Seção $\ref{sec:cp}$ são levantados conceitos básicos do artigo. Na Seção $\ref{sec:met}$, a metodologia proposta é apresentada. Em seguida, na Seção $\ref{sec:res}$, os resultados obtidos são descritos e discutidos. Finalmente, a Seção $\ref{sec:conc}$ apresenta conclusões e propostas de continuidade.

 \section{Conceitos Preliminares}
 \label{sec:cp}

\subsection{Modelos NARMAX}
Os modelos NARMAX (do inglês \textit{Non-linear AutoRegressive Moving Average model with eXogenous inputs}), \cite{CB1989}, são modelos discretos no tempo que explicam o valor da saída $y(k)$ em função de valores prévios dos sinais de saída e entrada. O modelo NARMAX polinomial pode ser representado como:
\begin{eqnarray}
 y(k) &=& F^l[y(k-1),\cdots ,y(k-n_y),  \nonumber\\
 &&u(k-1),  \cdots , u(k-d-n_u),   \\ 
&&e(k-1), \cdots , e(k-n_e)]+ e(k), \nonumber 
\end{eqnarray} 
em que $n_y$, $n_u$, $n_e$, $d$, $u(k)$ e $y(k)$ são os atrasos
em $y$, em $u$, em $e$, o tempo morto, o sinal de entrada
e o sinal de saída no instante $k$, respectivamente. $e(k)$ representa ruído e possíveis incertezas que não podem ser bem representados por $F^l$, que é uma função polinomial de $y(k)$, $u(k)$ e $e(k)$ com grau de não linearidade $l \in \mathbb{N}$.

Um possível subconjunto deste modelo é representado pelo modelo NARX ({\it Non-linear AutoRegressive model with eXogenous inputs}), o qual é matematicamente descrito por:
\begin{eqnarray}
 y(k) &=& F^l[y(k-1),\cdots ,y(k-n_y),  \\
 &&u(k-d),  \cdots , u(k-d-n_u)]+e(k) \nonumber. 
\end{eqnarray}

\subsection{Raiz do erro quadrado médio}
O RMSE pode ser calculado como segue \cite{aguirre2015}: 
\begin{equation}
RMSE= \frac{\sqrt{\sum_{N}^{k}(y(k)-\hat{y}(k))^2}}{\sqrt{\sum_{N}^{k}}(y(k)-\bar{y}(k))^2},
\label{eq:rmse}
\end{equation}
sendo $\hat{y}(k)$ a saída do modelo e $\bar{y}(k)$ a média da saída medida, $y(k)$.

\subsection{Aritmética intervalar}

Quando se trabalha com a aritmética intervalar, os números são representados por um limite inferior e um limite superior, obtendo assim um intervalo \cite{Ruetsch2005}. Intervalos são comumente representados por letras maiúsculas, tal como X. As extremidades de  X são denotadas frequentemente por \underline{X} e $\overline{X}$, respectivamente, de modo que X = [\underline{X}, $\overline{X}$]. Se suas extremidades são iguais \underline{X} = $\overline{X}$, esse novo número é um número real \cite{Rothwell2012}. 
A interseção  de dois intervalos $X \cap Y$ é um conjunto de números reais que pertence a ambos X e Y. A união $X \cup Y$ é um conjunto de números reais que pertence a X ou Y (ou ambos). Se $X \cap Y$ não é vazio, então $X \cap Y$ e $X \cup Y$ são intervalos que podem ser calculados por
\begin{eqnarray}
X \cap Y  = [\max(\underline{X}, \underline{Y}), \min(\overline{X}, \overline{Y})],\\
X \cup Y  = [\min(\underline{X}, \underline{Y}), \max(\overline{X}, \overline{Y})].
\end{eqnarray}  
As operações intervalares de adição, subtração e multiplicação são definidas como:
\begin{eqnarray}
X + Y = [\underline{X} + \underline{Y}, \overline{X} +  \overline{Y}],\\
X - Y = [\underline{X} - \overline{Y}, \overline{X} - \underline{Y}],\\
X \cdot Y = [\min\textit{(S)}, \max\textit{(S)}],
\end{eqnarray}
onde \textit{S} é o conjunto \{${\underline{X}\underline{Y},\underline{X}\overline{Y}, \overline{X}\underline{Y}, \overline{X}\overline{Y}}$\}. Se Y não contém o número zero, então o quociente X/Y é dado por
\begin{equation}
X/Y = X \cdot (1/Y) 
\end{equation}
onde $1/Y = [1/\overline{Y},1/\underline{Y}]$.

É importante mencionar que a adição e a multiplicação são associativas e comutativas, porém a distributividade não é geral para todos os casos, ou seja, duas expressões, que são equivalentes em aritmética verdadeira podem não ser equivalentes em aritmética intervalar \cite{Nepomuceno2016}.
Por exemplo, seja $X$ um intervalo,  a função $$ f(X)=X(1-X)$$ e uma extensão intervalar de $f$  $$ F_{1}(X)=X-X\cdot X .$$ Tem-se que $ f([0,1])=[0,1].(1-[0,1])=[0,1] $ enquanto $ F_{1}([0,1])=[0,1]-[0,1].[0,1]=[-1,1] $.

\begin{theo} \textbf{(Teorema Fundamental da Aritmética Intervalar \cite{Moore1979})}. Se $F$ tem propriedade de inclusão e é extensão intervalar de $f$, então

$f(X_1, \cdots, X_n) \subseteq F(X_1, \cdots, X_n)$.
\end{theo}
A prova e mais detalhes desse teorema podem ser encontradas em \cite{Moore1979}. O importante aqui é notar que a estimação de parâmetros pode ser vista como uma $f(\cdot)$ que é um subconjunto da realização intervalar proposta aqui, denotada então por $F(\cdot)$.

\subsection{Toolbox Intlab}

Computacionalmente pode-se utilizar o Intlab para cálculos envolvendo aritmética intervalar. O Intlab é um toolbox para Matlab que suporta cálculos com intervalos reais e complexos, vetores e matrizes.
Ele é projetado para ser rápido e possui desempenho compatível com outras rotinas pautadas apenas em ponto flutuante  \cite{Ru99a}.

A aritmética intervalar implica em computação com conjuntos. O Intlab foi projetado para obter rigorosas soluções. Uma solução para um determinado problema é produzida sob a forma de um intervalo que contêm a verdadeira solução. A aritmética finita da máquina é resolvida por meio de arredondamentos para o ponto flutuante mais próximo, isto é, arredondando a extremidade da esquerda para o número inferior da máquina mais próximo ou igual ao ponto final exato de um intervalo, e o ponto final direito para o número mais próximo da máquina maior do que ou igual ao exato ponto final direito \cite{Rothwell2012}.

O Intlab permite operações básicas a serem realizadas em intervalos reais e complexos, escalares, vetores e matrizes. Estas operações são inseridas semelhante a aritmética real e complexa no Matlab. Por exemplo, se a matriz $\mathbf{A}$ é introduzida, em seguida, $ A^2 $ realiza $\mathbf{A}\times \mathbf{A}$ em aritmética intervalar, enquanto que $ A.^2 $ resulta em cada componente de $\mathbf{A}$ elevado ao quadrado, usando aritmética intervalar \cite{Hargreaves02intervalanalysis}.

\section{Metodologia}
\label{sec:met}

Os sistemas experimentais foram excitados por um sinal PRBS, utilizando uma placa de aquisição de dados NI 6009 USB, da National Instruments.
Em seguida, as estruturas foram selecionadas utilizando a taxa de redução de erro juntamente com o critério de informação de Akaike \cite{Aka1974}, para cada um dos sistemas. 
Na etapa de determinação dos parâmetros, utilizou-se o método de mínimos quadrados clássico \cite{aguirre2015}.
Com o intuito de manusear os erros numéricos, utilizou-se o toolbox Intlab. Assim, durante a etapa de estimação dos parâmetros por mínimos quadrados intervalar os dados foram inseridos utilizando o comando \texttt{infsup}, de forma que os dados coletados pudessem ser computados como um intervalo. Desta forma, a inicialização do intervalo consiste no ínfimo e supremo computado, ou seja, o menor e o maior ponto flutuante próximo do dado inserido, tal que [a, b] esteja contido no intervalo c \cite{Ru99a}, \texttt{c = infsup(a,b)}.

Por exemplo, seja a = $0,25$ e b=$0,30$,

\noindent\texttt{infsup(a,b) = intval ans = 
[0.24999999999999,   0.30000000000000].    
}

É possível observar o cuidado do toolbox em arredondar para fora do intervalo, preservando o intervalo de interesse. Neste caso, o $0,25$ foi arredondado para baixo.

A partir disso, os mínimos quadrados intervalar utilizado, possui as incertezas numéricas incorporadas nos dados coletados,  a partir da resposta dos sistemas à excitação do PRBS. Assim, de posse dos dados intervalares estima-se os regressores intervalares.

Após a determinação dos parâmetros intervalares pelo método MQ intervalar, foi feita a validação do modelo obtido, utilizando dados diferentes da identificação. Por último, o índice RMSE foi computado, de modo a quantificar a qualidade dos modelos obtidos. Para cada caso, foi calculado o RMSE tradicional, por meio de simulação livre e por meio da validação um passo a frente e ainda o RMSE intervalar utilizando validação um passo a frente. Para o RMSE intervalar a saída do modelo $\hat{Y}(k)$ e a média da saída medida $\bar{Y}(k)$ são os intervalos obtidos.

  
 \section{Resultados}
 \label{sec:res}
  
 \subsection{Circuito RLC}
 
No primeiro estudo de caso, foi utilizado um circuito RLC série. 
Por meio da taxa de redução de erro juntamente com o critério de informação de Akaike, obteve-se a seguinte estrutura de um modelo NARX polinomial:
\begin{eqnarray}
\Psi&=&[y(k-1) \quad  y(k-1)y(k-2) \quad \nonumber \\
&&y(k-1)u(k-2) \quad u(k-1)].
\label{estrutura_narx}
\end{eqnarray} 
  
Os parâmetros determinados por meio dos Mínimos Quadrados intervalar são apresentados em (\ref{eq:13})
\begin{equation}
\centering
\small
\theta = \begin{bmatrix}
\scriptstyle [0,86273744299150; \quad    0,86526925655267] \\ 
\scriptstyle [-0,02018490110934; \quad   -0,01956475067962]\\ 
\scriptstyle [0,01184094798064;  \quad   0,01214790319626]\\
\scriptstyle [0,12910251755609,  \quad   0,13041151943168]
\label{eq:13}
\end{bmatrix}.
\end{equation}
 
\begin{figure}[H] 
		\centering
			\includegraphics[angle=0, scale=0.45]{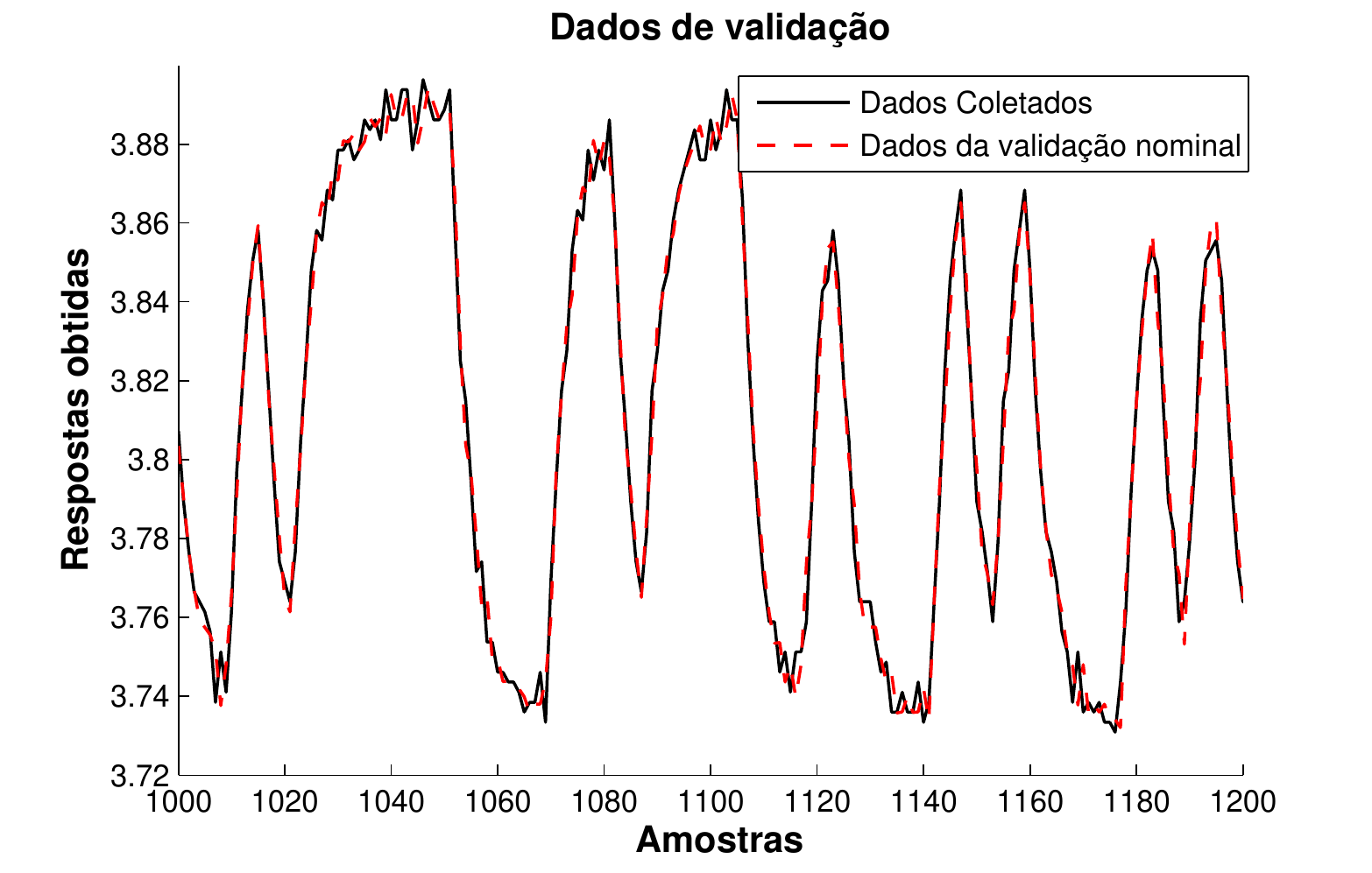} 
			\caption{Identificação do sistema RLC. (-) Dados coletados. (- -): Dados de validação do modelo identificado.} 
			\label{fig:1}
			\end{figure}

\begin{figure}[H] 
		\centering
			\includegraphics[angle=0, scale=0.42]{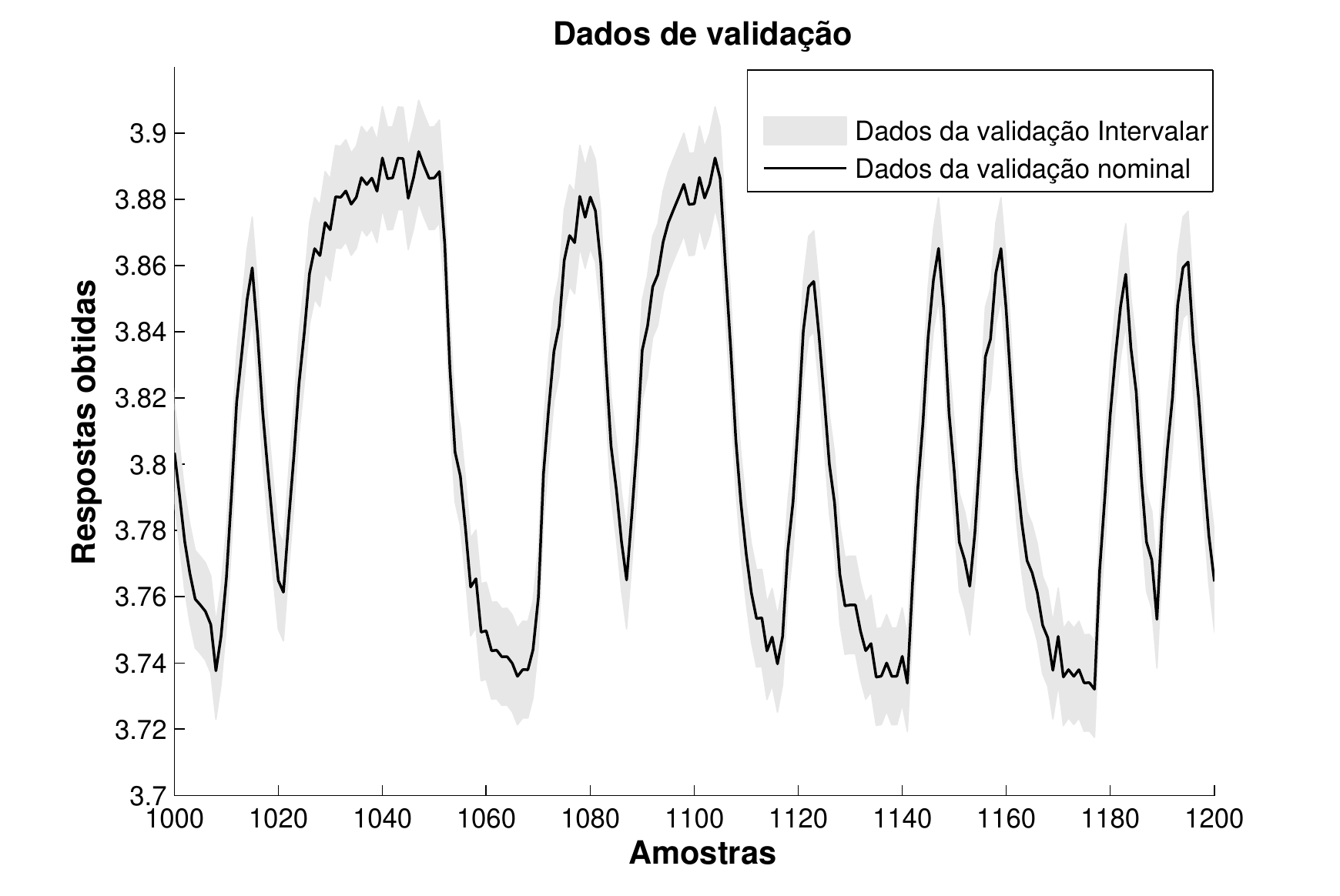} 
			\caption{Identificação Intervalar do sistema RLC. (-)  Dados de validação nominal do modelo identificado. (\crule{.3cm}{.3cm}) dados de validação intervalar do modelo identificado.} 
			\label{fig:2}
			\end{figure}

A Figura \ref{fig:1} apresenta a validação do modelo de forma tradicional para o circuito RLC, onde são apresentados os dados coletados juntamente com o modelo obtido, já a Figura \ref{fig:2} apresenta os dados de validação intervalar comparado com os dados de validação nominal.

\subsection{Sistema Eletromecânico}

\begin{figure}[ht!]
    \centering
    \includegraphics[scale=0.53]{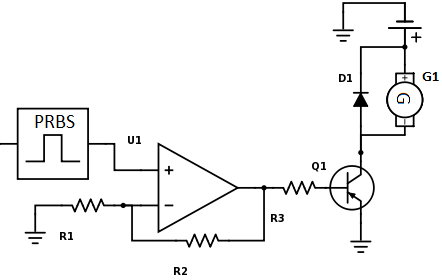}
    \caption{Esquema do sistema proposto.}
    \label{fig:motor}
\end{figure}

No segundo estudo de caso,  foi utilizado um sistema composto por 2 motores cc (6 V) acoplados, um motor e um funcionando como gerador. O esquema do circuito implementado é apresentado na
Figura \ref{fig:motor}. O sinal de entrada utilizado para excitar o sistema foi um sinal binário pseudoaleatório. 

 Por meio da taxa de redução de erro juntamente com o critério de informação de Akaike, a seguinte estrutura de modelo foi obtida:
\begin{eqnarray}
\Psi&=&[y(k-1) \quad  y(k-2) \quad  u(k-1) \nonumber \\ &&  u(k-1)y(k-1) \quad u(k-2)   \\ &&u(k-1)y(k-2) \quad   u(k-2)y(k-1) \nonumber \\&&  u(k-2)y(k-2) \quad y(k-2)^{2} \nonumber].
\label{eq:14}
\end{eqnarray} 
 
\begin{figure}[ht!] 
		\centering
			\includegraphics[angle=0, scale=0.5]{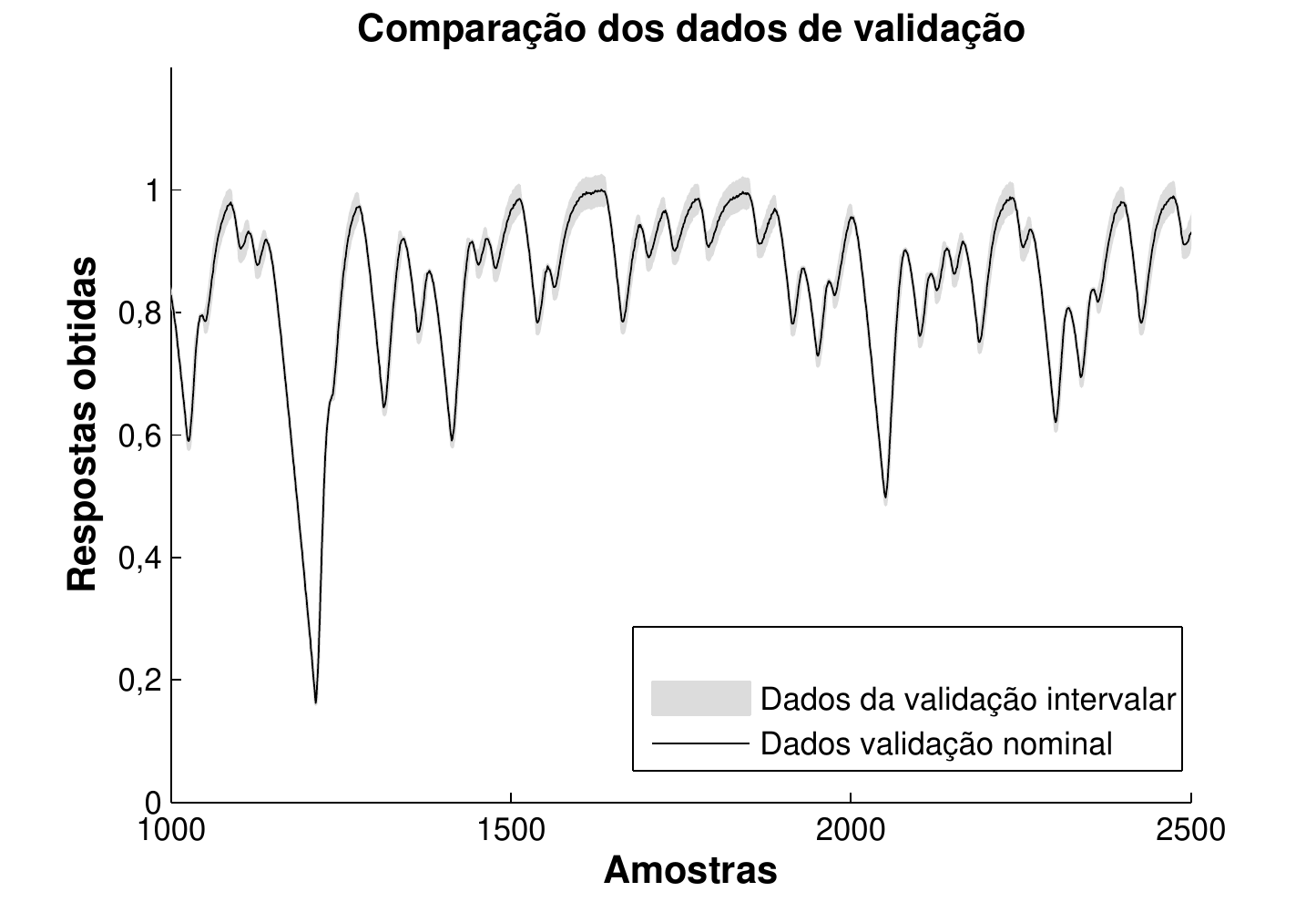} 
			\caption{Identificação Intervalar do sistema eletromecânico. (-)  Dados de validação nominal do modelo identificado. (\crule{.3cm}{.3cm}) dados de validação intervalar do modelo identificado.} 
			\label{fig:3}
			\end{figure}

Os parâmetros determinados por meio dos Mínimos Quadrados intervalar são apresentados em (\ref{eq:15})
\begin{equation}
\centering
\small
\theta = \begin{bmatrix}
\scriptstyle [1,86983289867678; \quad    1,87052216754597] \\
\scriptstyle[-0.87500271106274; \quad  -0,87433604737629] \\
\scriptstyle[   0,00877520297552; \quad    0,00879549049917] \\
\scriptstyle[  -0,07754127207970; \quad   -0,07522496598523] \\
\scriptstyle[   0,01156721443224; \quad    0,01159157820069] \\
\scriptstyle[   0,06736048557893; \quad    0,06969424138646] \\
\scriptstyle[  -0,09177956625574; \quad   -0,08882860964415] \\
\scriptstyle[   0,07814928825619; \quad    0,08111662045620] \\
\scriptstyle[   0,00273140158559; \quad    0,00275676329139] 
\label{eq:15}
\end{bmatrix}.
\end{equation}
A Figura \ref{fig:3} apresenta os dados de validação intervalar do modelo identificado juntamente com os dados coletados para o sistema eletromecânico.

\subsection{Oscilador \textit{Duffing-Ueda}}

O terceiro circuito analisado é o oscilador não-linear  Duffing-Ueda  que é definido como \cite{Billings2013}:
\begin{equation}
    \frac{d^2y}{dt^2}+k\frac{dy}{dt}+ \mu y^3=A\cos(t).
\end{equation}
Com amplitude A=1,2 e por meio da taxa de redução de erro juntamente com o critério de informação de Akaike, foi determinado os seguintes regressores: 
\begin{eqnarray}
\Psi&=& [y(k-1) \quad  y(k-2) \quad  y(k-3) \quad y(k-4) \nonumber \\ &&    y(k-5) \quad y(k-6) \quad   y(k-6)y(k-1)^2  \nonumber  \\&&   y(k-3)^3 \quad y(k-1)^3 \quad y(k-5)^3  \nonumber \\ &&  y(k-6)^3 \quad y(k-4)^3 \quad y(k-2)^3 \nonumber \\ &&  y(k-1)y(k-2)^2 \quad y(k-5)y(k-1)^2  \nonumber \\ && y(k-3)y(k-2)y(k-1) \nonumber \\&&  y(k-4)y(k-2)y(k-1)  \nonumber \\ &&  y(k-6)y(k-2)y(k-1)].
\label{eq:duff}
\end{eqnarray} 
 Os parâmetros determinados por meio dos Mínimos Quadrados intervalar são apresentados em (\ref{eq:18})
\begin{equation}
\small
\centering
\theta = \begin{bmatrix}
\scriptstyle[   1,90951882008706;  \quad 1,92646936612689] \\
\scriptstyle[   0,95365151698072;  \quad 1,00561820035704] \\
\scriptstyle[  -4,02822487143497;  \quad-3,98587426164288] \\
\scriptstyle[   1,44903214765727; \quad  1,46867281254767] \\
\scriptstyle[   1,48314782828422; \quad  1,52347135562097] \\
\scriptstyle[  -0,87118736563343; \quad -0,85672358123898] \\
\scriptstyle[   0,00225668759061; \quad  0,00236179315233] \\
\scriptstyle[   0,09300223610061; \quad  0,09376532378801] \\
\scriptstyle[  -0,06490883382016; \quad -0,06488422870018] \\
\scriptstyle[  -0,04798316559129; \quad -0,04727455439529] \\
\scriptstyle[  -0,00481684838248; \quad -0,00473039445345] \\
\scriptstyle[  -0,00453674339896; \quad -0,00386579675030] \\
\scriptstyle[  -0,01222583485781; \quad -0,01167347526245] \\
\scriptstyle[   0,03487125603343; \quad  0,03545208952191] \\
\scriptstyle[  -0,00540498279559; \quad -0,00528812953758] \\
\scriptstyle[  -0,00852990539515; \quad -0,00786729826986] \\
\scriptstyle[   0,00545335825399; \quad  0,00583508548234] \\
\scriptstyle[  -0,00118117654226; \quad -0,00107715645748] 
\label{eq:18}
\end{bmatrix}.
\end{equation}

\begin{figure}[ht!] 
		\centering
			\includegraphics[angle=0, scale=0.5]{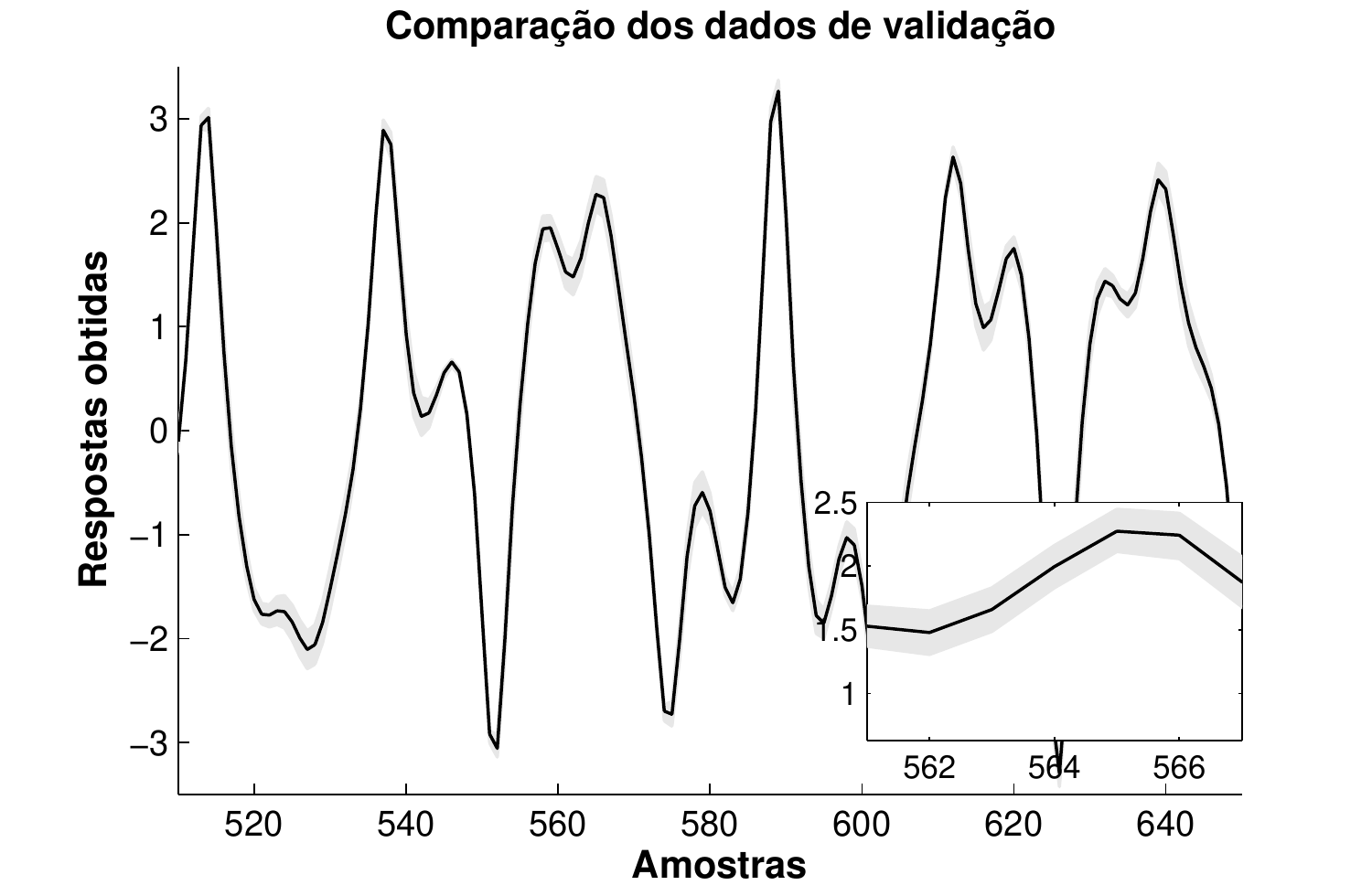} 
			\caption{Identificação Intervalar do oscilador Duffing-Ueda. (-) Em preto dados do sistema do modelo identificado. (\crule{.3cm}{.3cm}) dados de validação intervalar do modelo identificado.  A aproximação é para apresentar com clareza a largura do intervalo obtido nos dados de validação do sistema.} 
			\label{fig:4}
			\end{figure}

A Figura \ref{fig:4} apresenta os dados do sistema, juntamente com os dados de validação intervalar para o oscilador {\it Duffing-Ueda}.
Analisando as Figuras \ref{fig:2}, \ref{fig:3} e \ref{fig:4} é possível perceber que a resposta intervalar, obtida pelo Intlab, engloba as respostas obtidas de forma tradicional, conforme o esperado.

\subsection{RMSE}

A Tabela \ref{tab:1} apresenta o cálculo do RMSE para os três casos estudados, caso 1: circuito RLC, caso 2: sistema eletromecânico e caso 3: oscilador \textit{Duffing-Ueda}.  O RMSE foi calculado por simulação livre (RMSE 1) e por simulação um passo a frente (RMSE 2) e o RMSE intervalar foi obtido por meio da simulação um passo a frente, visto que por meio da simulação livre o intervalo cresce de forma indesejável. 
\begin{table}[ht!]
\centering
\small
\caption{Cálculo da raiz quadrada do erro quadrático médio (RMSE).}
\label{tab:1}
\begin{tabular}{c c cc}
\hline
Caso & RMSE 1 & RMSE 2 & RMSE Intervalar        \\ \hline
1    & 0,1099 &   0,1060       & [0,0040; \quad 0,3677] \\
2    & 0,1233  &   0,0079      & [0,0016; \quad 0,1030] \\
3    & 0,1776   &   0.0071     & [0,0003; \quad 0,0987] \\ \hline
\end{tabular}
\end{table}

\section{Conclusões}
 \label{sec:conc}

O presente artigo apresenta a identificação intervalar de três sistemas, um circuito RLC série, um sistema eletromecânico, composto por um motor e gerador e o oscilador {\it Duffing-Ueda}.  Foi simulado a taxa de redução de erro juntamente com o critério de Akaike, obtendo-se os modelos dos sistemas, por seguinte, por meio do \textit{toolbox} Intlab, simulou-se o algoritmo de mínimos quadrados intervalar para a etapa de identificação e validação dos sistemas, acrescentando apenas as incertezas numéricas.  

A principal contribuição deste artigo é apresentar a identificação intervalar de sistemas, que atribui uma robustez aos parâmetros determinados pelo método dos mínimos quadrados. Uma vez que sucessivos cálculos, acumulam erros e estes podem interferir na resposta do sistema analisado, principalmente ao trabalhar com inversão de matrizes. Assim, a  aritmética intervalar garante que a resposta verdadeira esteja contida dentro do intervalo obtido.

Pelo cálculo do RMSE, pode se concluir que os modelos foram obtidos de forma satisfatória. Entretanto, ao aplicar a aritmética intervalar analisando os intervalos obtidos para os regressores, pode-se afirmar que estes tem um largura considerável, o mesmo vale salientar para as respostas de validação que foram obtidas por meio de simulação um passo a frente, chegando uma variação em torno de milivolts.
Uma vez que, os erros numéricos vão se propagando a cada iteração não foi possível obter uma validação intervalar por simulação livre, ou seja, o intervalo cresceu de forma incontrolável. 

Em trabalhos futuros, pretende-se analisar formas de diminuir o intervalo na validação por simulação livre, a fim de obter uma maior contribuição para a identificação de sistemas.

\section*{Agradecimentos}
Agradecemos à CAPES, CNPq/INERGE, FAPEMIG e à Universidade Federal de São João del-Rei pelo apoio.


\bibliography{cba}
\end{document}